\documentstyle[epsf,12pt,cite]{article}

\begin{document}

\hyphenation{Ginz-burg Le-van-yuk  com-pound multilayered 
ap-pli-ca-tion con-sid-er com-pound su-per-con-duc-ting}

\newcommand{\ie}{{\it i.e.}}
\newcommand{\eg}{{\it e.g.}}
\newcommand{\etal}{{\it et al.}}

\newcommand{\eq}[1]{(\ref{#1})}

\newcommand{\YBCOf}{\mbox{Y$_1$Ba$_2$Cu$_3$O$_{7-\delta}$}}
\renewcommand{\epsilon}{\varepsilon}
\newcommand{\lsim}{\stackrel{<}{_\sim}}
\newcommand{\gsim}{\stackrel{>}{_\sim}}

\newcommand{\notacion}[2]{\newcommand{#1}{\mbox{$#2$}}}

\notacion{\Dsabe}{\Delta\sigma_{ab}(\epsilon)}
\notacion{\ns}{\left<\,n_s\,\right>}
\notacion{\nse}{\left<\,n_s(\epsilon)\,\right>}
\notacion{\DM}{\Delta M}
\notacion{\DMe}{\Delta M (\epsilon)}
\notacion{\Ds}{\Delta\sigma}
\notacion{\Dse}{\Delta\sigma(\epsilon)}
\notacion{\Tc}{T_C}
\notacion{\kB}{k_{\rm B}}
\notacion{\DMeh}{\Delta M(\epsilon)_{\mbox{\it h}}}
\notacion{\Tsuper}{T^C}
\notacion{\epsilonsuper}{\epsilon^C}

\title{\hrule \mbox{}\\ 
\centerline{\large\rm Published in: \sc Europhys.~Lett., {\bf 59} \rm
(5), pp. 754-760 (2002).}\vspace{-0.3cm} \mbox{}\\ \hrule \mbox{}\\
\mbox{}\\ \mbox{}\\
\Large\bf
On the consequences of the
uncertainty principle\\
on the
superconducting fluctuations\\
 well inside the normal state\\ \mbox{}\\ 
}

\author{\normalsize 
F\'elix Vidal, Carlos Carballeira,
Severiano~R.~Curr\'as, 
Jes\'us Mosqueira, \\
\normalsize Manuel
V.~Ramallo, Jos\'e Antonio Veira, and Jos\'e Vi\~na.
\\ \mbox{} \\ \normalsize
Laboratorio de Baixas Temperaturas e
Superconductividade,\\ \normalsize Departamento de F\'{\i}sica da Materia
Condensada,\\ \normalsize Universidade de Santiago de Compostela,\\
\normalsize Santiago de Compostela E-15782, Spain. }

\date{}

\maketitle

\begin{abstract}
We first argue that the  collective behaviour  of the
Cooper pairs created by thermal fluctuations well  above the
superconducting transition temperature, \Tc, is dominated by the
uncertainty principle which, in particular, leads to a well-defined
temperature, \Tsuper, above which the superconducting coherence
vanishes. On the grounds of the BCS approach, the corresponding
reduced-temperature,
$\epsilonsuper\equiv\ln(\Tsuper/\Tc)$, is estimated to be around
$0.55$,
\ie, above $\Tsuper\simeq1.7\,\Tc$ coherent Cooper pairs cannot
exist. The implications of these proposals on the superfluid
density are then examined using the Gaussian-Ginzburg-Landau
approximation. Then we present new measurements of the thermal
fluctuation effects on the electrical conductivity and
on the magnetization in different low-  and high-\Tc\
superconductors with different dopings which are in excellent
agreement with these proposals and that demonstrate the
universality of \epsilonsuper.
\\ \mbox{} \\ \\ \mbox{} \\ \footnotesize
74.20.-z\ \ Theories and models of superconducting state\\
\footnotesize 74.20.De\ Phenomenological theories
(two-fluid, Ginzburg-Landau, etc.)\\ \footnotesize
74.40.+k~Fluctuations (noise, chaos, nonequilibrium
superconductivity, localization, etc.) 
\end{abstract}

\newpage

The behaviour of the superconducting fluctuations well inside the
normal state, when the superconducting coherence length,
$\xi(T)$, becomes of the order of its amplitude extrapolated
at $T=0$~K, $\xi(0)$, is a long-standing open problem which
interest has been considerably enhanced by the discovery of the
high temperature cuprate
superconductors.\cite{ns,Tinkham,altaTc,bajaTc,micro,reviewScience}
Recent measurements of the superconducting fluctuations above
\Tc\ in various high-\Tc\cite{altaTc} and  low-\Tc\cite{bajaTc}
superconductors suggest the existence  of a well-defined
reduced-temperature,
$\epsilonsuper\equiv\ln(\Tsuper/\Tc)$, above which the fluctuation
effects vanish. These results were explained  phenomenologically  by
introducing in the Gaussian-Ginzburg-Landau (GGL) approach  a
so-called ``total-energy''
cutoff,\cite{altaTc,bajaTc} instead of the
conventional momentum cutoff always
used until now.\cite{ns,Tinkham,micro} The first aim of
this Letter is to analyze the physical origin of such a total-energy
cutoff. We will see here that it can be easily understood in terms
of the uncertainty principle, which imposes a limit to the shrinkage
above \Tc\ of the superconducting wave function when the temperature
increases. Then, we will probe
experimentally the universality of \epsilonsuper\ by 
extending our previous measurements of the fluctuation effects in the
high reduced-temperature region to very different low- and high-\Tc\
superconductors with different dopings, including for the first time
a type-I superconductor.

The Heisenberg uncertainty principle was applied  in
 1953 to superconductors by Pippard to relate
the  size  of the ``wave packet formed by the electronic
states'', $\xi_0$, to  the normal-superconducting transition
 temperature, \Tc~\cite{Pippard,deGennes}. In terms of the
BCS approach, the Pippard proposal suggests that the minimum size
of a Cooper pair is of the order of $\xi_0$. By taking into account
that the superconducting coherence length,
$\xi(T)$, is the characteristic distance over which the density of
Cooper pairs may vary, it may be concluded that {\it even above
\Tc}, where the Cooper pairs are created by thermal fluctuations,
$\xi(T)$ must verify
\begin{equation}
\xi(T) \gsim \; \xi_0,
\label{cutoffgeneral}
\end{equation}
where $\xi_0$ is the actual superconducting coherence length at
$T=0$~K. Equation (\ref{cutoffgeneral})
provides then a  constraint for the existence and behaviour of
coherent Cooper pairs which  must apply to any theoretical
description of the superconducting state formation, including those
non-BCS-like which are being proposed for  cuprate
superconductors~\cite{reviewScience}. In fact, this condition may be
directly obtained by applying the uncertainty principle to the
Cooper pairs localized in a coherent volume. In other words,
Eq.~(\ref{cutoffgeneral})  accounts for  the  limits imposed by the
uncertainty principle to the shrinkage, when the temperature
increases, of the superconducting wave function. 

The most outstanding consequence of Eq.~(\ref{cutoffgeneral}) is that
it  naturally leads  to a reduced-temperature,
\epsilonsuper, determined by
$\xi(\epsilonsuper)\simeq\xi_0$, above which  coherent Cooper pairs
cannot exist. The value of
\epsilonsuper\ will depend on each particular approach through  the
$\epsilon$-dependence of
$\xi(\epsilon)$, where $\epsilon\equiv\ln(T/\Tc)$ is the
reduced-temperature. For
instance, by using the mean-field reduced-temperature dependence 
of the coherence length\cite{validoGGL},
$\xi(\epsilon)=\xi(0)\epsilon^{-1/2}$, then
$\epsilonsuper\simeq(\xi(0)/\xi_0)^2$. On the grounds of the BCS
approach in the clean limit, $\xi(0)=0.74\, \xi_0$, and so
$\epsilonsuper\simeq0.55$ in these superconductors. In fact, this
striking result probably also holds at a qualitative level in
moderately  dirty BCS superconductors  (when $\ell\lsim\xi_0$, $\ell$
being the mean free path of the normal carriers) because one may
expect  that both the Ginzburg-Landau (GL) coherence length
amplitude and the actual superconducting coherence length at $T=0$~K
will be affected by impurities to a similar extent (see,
\eg, Ref.~\cite{deGennes}, \S\,7-2).

What consequences has Eq.~(\ref{cutoffgeneral}) on the
collective behaviour of the Cooper pairs  at reduced-temperatures
below
\epsilonsuper?  Probably the simplest way to address this question
at a qualitative level is to introduce Eq.~(\ref{cutoffgeneral}) in
the conventional GGL framework, in spite that the latter
formally applies only when $\epsilon\ll 1$.\cite{validoGGL} In
fact, another central point in this Letter will be to probe
experimentally if the introduction of Eq.~(\ref{cutoffgeneral}) in
the GGL approximation extends its applicability up to \epsilonsuper.
On the grounds of the GGL approach, Eq.~(\ref{cutoffgeneral}) leads
to
\begin{equation}
\xi^{-2}(\epsilon)+
k^2 \; \lsim \;  \xi_0^{-2},
\label{cutoffGGL}
\end{equation}
where the left-hand term is the GGL {\it
total\/} energy per superconducting carrier (in units of
$\hbar^2/2m^*$, where $m^*$ is the superconducting carriers 
effective mass) of the fluctuating mode with wave vector {\bf k}.
Actually, as stressed before for Eq.~(\ref{cutoffgeneral}), one
may directly obtain Eq.~(\ref{cutoffGGL}) by na\"{\i}vely applying
the uncertainty principle to the spatial extension of the
superconducting fluctuations  and taking into account  that at 
finite temperatures the energy balance when creating the 
fluctuations must include  both the uncertainty principle and the
thermal agitation. This crude reasoning is to some extent similar to
the well-known textbook procedure used to estimate the minimum size
of atoms by balancing the Coulomb and the Heisenberg localization
energies.\cite{Feynman}  Note  that for
$\epsilon\ll\epsilonsuper$ [and using
$\xi(0)$ instead of $\xi_0$] Eq.~(\ref{cutoffGGL}) reduces  to the
widely used kinetic-energy or momentum cutoff condition
$k^2\leq c\,\xi^{-2}(0)$, where $c$ is a cutoff constant of the
order of unity~\cite{ns,Tinkham,micro}.   Note also
that any cutoff condition is not built-in in the standard GGL
equations, and it must be added to them ``by hand''. Adequate
extensions of the standard  GGL equations to further Gor'kov
perturbative orders could perhaps reproduce the total-energy cutoff,
but we are not aware of any successful attempt of this.

The  deep influence of Eq.~(\ref{cutoffGGL}) on the
superconducting fluctuations in the normal state below 
\epsilonsuper\ is well illustrated by the reduced-temperature
dependence  of the
 superfluid  density,
\nse, on which any observable will
depend.    \nse\ is defined here on
the grounds of the conventional GL approach  as  the spatially- and
thermally-averaged squared modulus of the 
GL wave
function~\cite{ns,Tinkham}:
\begin{equation}
\nse\equiv\left<\left|\Psi\right|^2\right>
=\sum_{\bf k}
\left<\left|\Psi_{\bf k}\right|^2\right>,
\label{defns}
\end{equation}
where $\Psi_{\bf k}$ is the wave function of the fluctuating mode
with wave vector ${\bf k}$. On the grounds of the Gaussian
approximation for the fluctuations of both the amplitude and phase
of the order parameter, and for zero applied magnetic field, 
$<\left|\Psi_{\bf k}\right|^2>$  is given by
the well-known  result (see, \eg, Eq.~(8.27) of
Ref.~\cite{Tinkham}):
\begin{equation}
\left<\left|\Psi_{\bf k}\right|^2\right>=
{{2m^*\xi^2(0)}\over{\hbar^2}}\,
{{\kB T}\over{\epsilon+k^2\xi^2(0)}}. 
\label{modos}
\end{equation}
This  familiar expression of the
amplitude average of each fluctuating mode already illustrates
the need of a total-energy cutoff: 
$<\left|\Psi_{\bf k}\right|^2>$ increases (instead of
decreasing) when the temperature increases well above \Tc. This
nonphysical behaviour appears for {\it any} $k$-value at 
reduced-temperatures  above 
$\epsilon+k^2\xi^2(0)=1$.  It cannot be eliminated, therefore,
by the momentum cutoff condition.
In contrast, a total-energy cutoff
will remove, {\it if} $\epsilonsuper\leq 1$, this nonphysical
behaviour  (see also later). 

From Eqs.~(\ref{defns}) and (\ref{modos}) the superfluid density for
any cutoff criterion may be directly calculated by just imposing 
on the
${\bf k}$-summation the corresponding limits for the modulus of
${\bf k}$. Such a limit  is  infinity for no cutoff, 
$\sqrt{(\epsilonsuper-\epsilon)}/\xi(0)$ for the total-energy
cutoff, and $\sqrt{c}/\xi(0)$ for the momentum cutoff. As it is well
known, without cutoff the above integrations diverge at every
temperature~\cite{ns,Tinkham}. With the  total-energy cutoff, we
get for 2D-films of thickness $d$ (and also for extremely anisotropic
layered superconductors with effective interlayer separation $d$):
\begin{equation}
\nse_E^{\rm 2D}={{m^*\kB T}\over{2\pi\hbar^2d}}
\ln\left(
{{\epsilonsuper}\over{\epsilon}}
\right),
\label{nsE2D}
\end{equation}
and for 3D-bulk isotropic superconductors:
\begin{equation}
\nse_E^{\rm 3D}={{m^*\kB
T}\sqrt{\epsilon}\over{\pi^2\hbar^2\xi(0)}}
\left(
\sqrt{{{\epsilonsuper-\epsilon}\over{\epsilon}}}
\;-\;
\arctan
\sqrt{{{\epsilonsuper-\epsilon}\over{\epsilon}}}
\right).
\label{nsE3D}
\end{equation}
The corresponding
expressions under a momentum cutoff may be obtained from
Eqs.~(\ref{nsE2D}) and (\ref{nsE3D}) by replacing $\epsilonsuper$ by
$c+\epsilon$. 

The reduced-temperature dependence of
the superfluid density under these different cutoff conditions is
shown in Fig.~1(a).  We have chosen for representation 
$\epsilonsuper=0.55$   as it results from the simplest estimate  using
the BCS theory in the clean limit (or, in the case of the momentum
cutoff, $c=0.55$).  All the other parameters entering in
Eqs.~(\ref{nsE2D}) and (\ref{nsE3D}), including $\xi(0)$,
are absorbed by the normalization chosen in the plot. As clearly
illustrated by the figure,  the conventional momentum cutoff
predicts   a nonphysical increase of the superfluid density  at high
$\epsilon$'s, which is a consequence of the high-$T$ divergence  
of Eq.~(\ref{modos}).
 In contrast, under the total-energy cutoff \nse\ vanishes
at $\epsilon=\epsilonsuper$. The corresponding 
fall-off
  is remarkably sharp, following in the close vicinity of 
\epsilonsuper\ a  
 power law-like behaviour with respect  to $|\tilde\epsilon|$, with
$\tilde\epsilon\equiv\ln(T/\Tsuper)=\epsilon-\epsilonsuper$: In the
2D superconductors 
$\nse_E^{2D}\stackrel{\propto}{_\sim}|\tilde\epsilon|$, and in the
3D superconductors 
$\nse_E^{3D}\stackrel{\propto}{_\sim}|\tilde\epsilon|^{1.5}$,  in
both cases with  5\% or better accuracy for
$|\tilde\epsilon|\leq0.1$ if  
$\epsilonsuper\geq0.3$. 
Also  noticeable is that such a rapid fall-off occurs  when the
coherence length competes with the  size of
the individual Cooper pairs, as  illustrated by
the  
$\xi(\epsilon)/\xi_0$ scale shown in Fig.~1(a) (whose   numerical
values correspond to the BCS clean limit): When
$\xi(\epsilon)\gsim2\xi_0$, which  roughly corresponds to
$\epsilon\lsim0.15$, the consequences of the uncertainty principle
on the superconducting fluctuations are inappreciable, and the
momentum cutoff condition provides a good approximation to \nse. 
However, for
$2\xi_0\gsim\xi(\epsilon)\gsim\xi_0$ (corresponding to
$0.15\lsim\epsilon\lsim\epsilonsuper$), \ie, when
$\xi(\epsilon)$ competes with the  size of the individual Cooper pairs,
the uncertainty principle will dominate the collective behaviour of
these Cooper pairs.

The above results strongly suggest,
therefore, that any  
superconducting fluctuation effect in the normal state will vanish
above
\epsilonsuper.
 However, probably the best way to probe  
these results is to extend our
first experiments on the  fluctuation effects in the
high-$\epsilon$
region~\cite{altaTc,bajaTc} to
different superconducting materials with  different
$\xi(0)$ values and also with various pairing states and maybe
different pairing mechanisms. 
Some
examples of the fluctuation effects on the magnetization [the
so-called fluctuation magnetization, \DM] and on the
electrical conductivity [the so-called paraconductivity,
$\Ds(\epsilon)$]  obtained in different low- and high-\Tc\
superconductors are presented in Fig.~1(b). The experimental setups
and procedures used in our present experiments are similar to those
described in Refs.~\cite{altaTc} and
\cite{bajaTc}. Other aspects, including the preparation of the
samples, are going to be published elsewhere. Let us stress here,
however, that the samples used in the paraconductivity experiments
were single crystals and epitaxial thin films. The electrical
resistivity versus temperature was measured with a four-terminal
arrangement by using a conventional low-frequency (37 Hz) ac lock-in
amplifier phase sensitive technique. To be able to determine \DM\ in
the high-$\epsilon$ region, the magnetization measurements were
performed with a SQUID magnetometer (Quantum Design, model MPMS) and
by using quite big polycrystalline samples (with masses up to a few
grams). Let us also note here that in analyzing all these measurements 
the background fitting
 region was always localized well above
$\epsilonsuper$. 
So, the results presented in
Fig.~1(b) are only moderately affected by the background
uncertainties. 
These measurements cover
almost two orders of magnitude in reduced   magnetic fields
$h\equiv H/H_{c2}(0)$, where
$H_{c2}(0)$ is the upper critical  
magnetic field  amplitude (extrapolated to $T=0$~K). In fact, they
cover both the zero  field  limit ($h\ll\epsilon$) and  the
finite  field regime. In this last case the \DMeh\ data could be
also affected by dynamic and non-local electrodynamic
effects~\cite{ns,Tinkham}. These examples also cover quite
different superconductors,  including  moderately  dirty
(PbIn8\%, \ie, Pb-8$\,$at.$\,$\%$\,$In)  and clean (all the others),
type~I (Pb) and II (all the others), 3D-bulk low-\Tc\ 
(isotropic PbIn8\% and Pb; moderately anisotropic MgB$_2$) and
2D-layered high-\Tc\  (the optimally-doped Bi-2212,
\ie, Bi$_2$Sr$_2$CaCu$_2$O$_{8+\delta}$ with $\delta\simeq0.16$, 
 and in the inset the underdoped
LaSCO/0.1, \ie, La$_{1.9}$Sr$_{0.1}$CuO$_4$). 

The hallmark of the
suppression of the superfluid  density by the total-energy cutoff,
\ie, the rapid fall-off at a 
well-defined reduced-temperature, $\epsilonsuper$, of the
superconducting fluctuation effects, is present in all the
experimental curves in Fig.~1(b).  The dashed and the
solid lines in the main Fig.~1(b) and in its inset correspond to the GGL
predictions in the zero-field limit under a conventional momentum
cutoff and, respectively, a total-energy cutoff. The
corresponding formulas may be found in
Refs.~\cite{altaTc,bajaTc}. Note that for our present purposes 
we must substitute  $c$ by
\epsilonsuper\ in the expressions under the total-energy cutoff, and
that in the case of
the 2D-multilayered superconductors studied here we must
multiply the corresponding single-layered expressions by $N$, the
number of superconducting layers per periodicity
length. The
$\xi(\epsilon)/\xi_0$ scales in Fig.~1(b) and in its inset  illustrate
that the behaviour of the superconducting fluctuations is dominated
by the localization energy when the superconducting coherence 
competes with
$\xi_0$, the  size of the Cooper pairs. The comparison between the two
\DMeh-curves for PbIn8\%  in Fig.~1(b) shows that the data for $h=0.1$
are, below $\epsilon\simeq0.1$, appreciably affected by finite-field
(or Prange) effects, but their \epsilonsuper\ still remains unchanged
well within the experimental uncertainties. A more detailed comparison
with the theoretical results in this finite-field regime will be
published elsewhere.

The \epsilonsuper-values for all the compounds studied here,
included those of Fig.~1(b) and also the ones measured in
Refs.~\cite{altaTc,bajaTc},  are
presented in Fig.~2 as a function  of
$\xi(0)$. In this figure, Tl-2223  and Y-123 stand for the
optimally-doped Tl$_2$Ba$_2$Ca$_2$Cu$_3$O$_{10}$ and
Y$_1$Ba$_2$Cu$_3$O$_{7-\delta}$ with $\delta\simeq0.05$, 
 LaSCO/0.14
and 0.25 stand for  the optimally-doped
La$_{1.86}$Sr$_{0.14}$CuO$_4$ and, respectively, overdoped
La$_{1.75}$Sr$_{0.25}$CuO$_4$, and PbIn18\% stands for the
Pb-18at.\%In alloy.  As it
may be clearly seen,
\epsilonsuper\  varies less than  a factor 3,  while these
measurements  cover almost two orders of magnitude  in
 coherence lengths, in amplitudes of the
fluctuation effects at 
$\epsilon=0.01$ (on both
\DM\ and \Ds), or (in the case of \DM) in applied reduced magnetic
fields  [$h\equiv H/H_{c2}(0)$ covers in these experiments
the range
$2\times 10^{-3} \lsim h \lsim 2 \times
10^{-1}$]. Therefore, these
results  provide  strong experimental evidence that  the
suppression  of the
superconducting
coherence above a well-defined
temperature
\Tsuper\ in the normal state is
 due to a universal mechanism. These results do not exclude,
indeed, a possible variation of \epsilonsuper\ in extremely dirty
superconductors or in the high reduced magnetic field region (when
$h\rightarrow 1$), where different non-local and
pair-breaking effects could appear~\cite{ns,Tinkham}.

In conclusion, the experimental results and the  analyses 
presented here  suggest that the behaviour of the superconducting
fluctuations at high reduced-temperatures is dominated by the
uncertainty principle, which imposes a limit to the shrinkage of
the superconducting wave function. These ideas provide a physical
meaning to the ``total-energy'' cutoff heuristically introduced 
before\cite{altaTc,bajaTc} to extend the applicability of the
mean-field--like approximations to
describe the superconducting fluctuations above
\Tc\ from $\epsilon\ll 1$ up to \epsilonsuper. It would be also
interesting to extend these analyses to other superconductors, very in
particular to the heavy fermions whose superconducting state coupling
is based on antiferromagnetic spin fluctuations. Our present results  
may also have implications  beyond the superconducting fluctuations
issue. For instance, the striking fact that even in cuprates with
different dopings (including  the underdoped)
\epsilonsuper\ takes the value
  that  we have directly obtained by
using the  BCS relationship between 
$\xi(0)$ and $\xi_0$ provides a constraint  to any theoretical
description of the superconducting state formation in these
compounds,  whose implications will deserve further analysis. The
possible implications of these results on the so-called
zero-dimensional superconductors\cite{ns,Tinkham}, when $\xi(T)$
becomes bigger than the dimensions of the material in all directions,
will also deserve further analysis. In fact, our present results
suggest that the uncertainty principle constraint on the
superconducting wave function will provide the last limit to the
smallness of an {\it isolated} superconductor. They also suggest that
the uncertainty principle must be taken into account when describing the
short-wavelength thermal fluctuations around  any phase transition with
a quantum order parameter, so that the classical cutoff condition,
$k\lsim\xi^{-1}(0)$, must be substituted by a total-energy cutoff
which takes  into account the shrinkage at high reduced-temperatures
of the quantum wave function. 

\mbox{}

\nonfrenchspacing
This work is dedicated to the memory of Victoriano Reinoso, President
of Uni\'on Fenosa, whose support through a grant of Uni\'on Fenosa
(No.~0666-98) is gratefully acknowledged. Some of the data points
summarized in Fig.~2 were obtained in samples supplied by
J.A.~Camp\'a, A.~Maignan, I.~Rasines, A.~Revcolevschi, and P.~Wagner.

\mbox{}

\newpage


\begin{figure}[ht]
\mbox{}\vspace{0.5cm}\mbox{}\\
\epsfxsize=17cm
\epsfbox{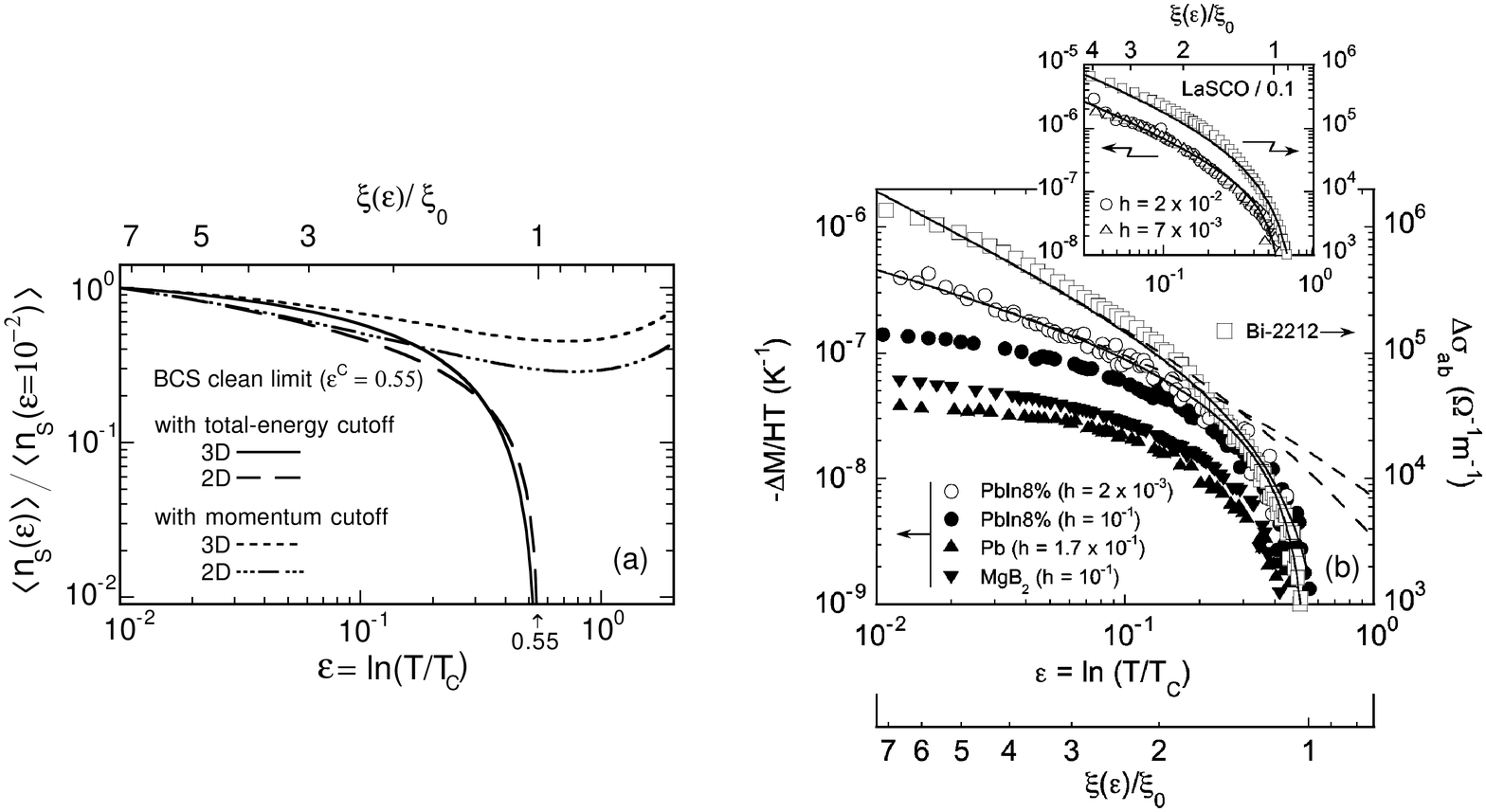}
\caption{
(a) Reduced-temperature dependence of 
the superfluid density above \Tc, 
calculated on the grounds of the  GGL approach for zero
applied magnetic field and in the 3D and 2D limits under both the 
momentum  and the total-energy cutoffs. (b) Some  examples of the
reduced-temperature dependence of the  fluctuation-induced
magnetization    in the zero field limit (open symbols) and in the
finite field regime (solid symbols), and of the fluctuation-induced
 electrical conductivity, measured in different superconductors. In
the anisotropic superconductors, the magnetic field is applied
perpendicular to the in-plane directions. The dashed and solid
lines in the panel (b) and its inset correspond to the GGL
calculations in the zero field limit under a momentum cutoff and,
respectively, a total-energy cutoff. The
$\xi(\epsilon)/\xi_0$ scales in both (a) and (b), whose values
correspond to the BCS clean limit, illustrate that the behaviour of
superconducting fluctuations is dominated by the uncertainty principle
when
$\xi(\epsilon)$ competes with $\xi_0$, the size of the Cooper pairs. }
\end{figure}

\newpage


\begin{figure}[ht]
\mbox{}\vspace{-0.0cm}\mbox{}\\
\epsfxsize=11cm
\centerline{\epsfbox{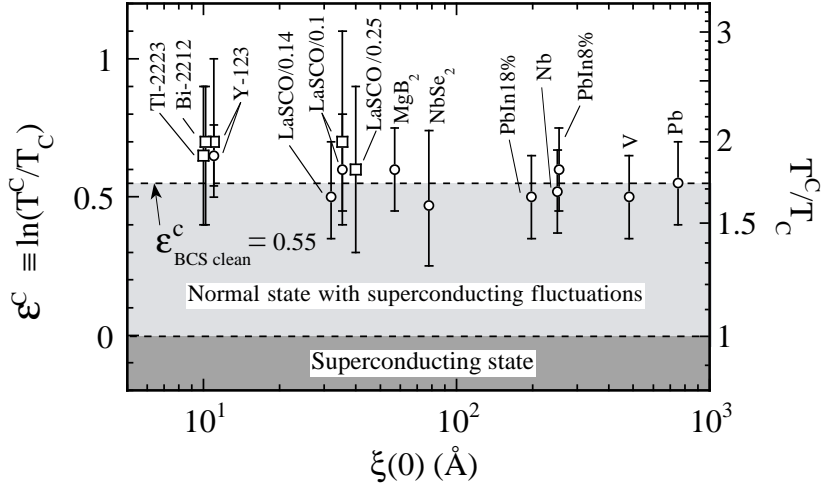}}
\mbox{}\\
\caption{
Values of
\epsilonsuper, the reduced-temperature at which the superconducting
fluctuations above \Tc\ vanish,  obtained from measurements of the
thermal fluctuation effects on the magnetization (circles) or on the
electrical conductivity (squares) in different superconducting
materials, as a function of their Ginzburg-Landau coherence length
amplitude
$\xi(0)$. The data points for Tl-2223, Bi-2212, Y-123, MgB$_2$, PbIn18\%
and PbIn8\% were taken from Refs.\cite{altaTc} and \cite{bajaTc}. In the
case of the anisotropic superconductors,
$\xi(0)$ 
 corresponds to the in-plane
coherence length amplitude. The
magnetization data always correspond  to
$h\leq0.2$.  The  error bars represent well all the experimental
uncertainties, included those  associated with the background
estimation. The line
$\epsilonsuper=0.55$  corresponds to the
simplest theoretical estimate using the BCS model in  the clean 
 limit. 
}
\end{figure}


\begin{thebibliography}
\bibitem{micro}{Previous works on the high-$\epsilon$ behaviour of the
paraconductivity in LTSC and HTSC have been quoted in
Refs.\cite{altaTc} and
\cite{bajaTc}.}
\end{thebibliography}
\end{document}